\documentclass[nofootinbib]{revtex4}
\usepackage{amsmath}
\usepackage{amssymb}

\begin{document}

\title{\Large
New Forms of BRST Symmetry on a Prototypical First-Class System
}

\author{Bhabani Prasad Mandal\footnote {e-mail address: bhabani.mandal@gmail.com}}

\affiliation{Department of Physics,
Banaras Hindu University,\\ Varanasi - 221005, India.}
 
\author{Sumit Kumar Rai\footnote {e-mail address: sumitssc@gmail.com}}

\affiliation{Sardar Vallabhbhai Patel College,\\
Veer Kunwar Singh University,\\Ara, Bhabua - 821101, India.}

\author{Ronaldo Thibes\footnote{e-mail address: thibes@uesb.edu.br}}
\affiliation{Departamento de Ci\^encias Exatas e Naturais,\\
Universidade Estadual do Sudoeste da Bahia,
Itapetinga - 45700000, Brazil.}

\begin{abstract}
We scrutinize the many known forms of BRST symmetries, as well as some new ones, realized within a prototypical first-class system.  Similarities and differences among ordinary BRST, anti-BRST, dual-BRST and anti-dual-BRST symmetries are highlighted and discussed.  We identify a precise $\mathbb{Z}_4\times\mathbb{Z}_2$ discrete group of symmetries of the ghost sector, responsible for connecting the various forms of BRST transformations.  Considering a Hamiltonian approach, those symmetries can be interrelated by canonical transformations among ghost variables.  However, the distinguished characteristic role of the dual BRST symmetries can be fully appreciated within a gauge-fixed Lagrangian viewpoint.  New forms of BRST symmetries are given, a set generalizing particular ones previously reported in the literature as well as a brand new unprecedented set.   The featured gauge invariant prototypical first-class system encompasses an extensive class of physical models and sheds light on previous controversies in the current quantum field theory literature.
\end{abstract}

\maketitle

\section{Introduction}
The fundamental  Becchi-Rouet-Stora-Tyutin (BRST) transformation mixing physical fields with the anticommuting Faddeev-Popov ghosts \cite{Faddeev:1967fc} has been around since the early seminal works of Becchi, Rouet, Stora and Tyutin  
\cite{Becchi:1974xu, Becchi:1974md, Tyutin:1975qk, Becchi:1975nq}, when it was originally introduced in the context of Yang-Mills theories to assure renormalizability and unitarity.  It is one of the most prominent symmetries in physics, which remains at quantum level even after the introduction of explicit gauge symmetry breaking terms in the classical Lagrangian for calculation purposes.  In fact, that persisting odd symmetry was first used in a Lagrangian context to show that the existence of Slavnov identites at all orders of perturbation theory guaranteed the renormalization of the Abelian Higgs-Kible model  \cite{Becchi:1974xu, Becchi:1974md}, with a subsequent generalization to non Abelian gauge field theories in the sequel paper \cite{Becchi:1975nq}.  By around the same time, the closely related Hamiltonian Batalin-Fradkin-Vilkovisky (BFV) approach was under construction \cite{Fradkin:1973wke, Fradkin:1975cq, Batalin:1977pb, Fradkin:1977xi}.  Aiming to work with relativistic gauges in the quantization of Hamiltonian constrained systems achieving both covariance and unitarity \cite{Fradkin:1973wke}, the BFV quantization scheme \cite{Fradkin:1975cq, Batalin:1977pb, Fradkin:1977xi} augmented the original phase space also including anticommuting ghost fields -- realizing the BRST symmetry as canonical transformations in that extended phase space in a gauge-independent way.   It became immediately clear that those two Lagrangian and Hamiltonian approaches could be fully connected by a judicious choice of gauge, aligned with specific momenta integrations in the generating functional. The BFV quantization method has proved to be quite general and, as an important bonus, works perfectly fine for systems with open or reducible gauge algebras as well.
Besides that original key BRST symmetry, an akin interwoven one, with ghosts and anti-ghosts interchanging roles, was soon reported in the literature  \cite{Curci:1976bt, Curci:1976ar, Ojima:1980da} and came to be known as {\it anti}-BRST symmetry \cite{Hwang:1983sm, Baulieu:1983tg, Hwang:1989mn}.
That pair of graded symmetries -- BRST and anti-BRST -- can be jointly inserted in a quantum Hamiltonian framework for general gauge invariant systems  \cite{Gregoire:1991rs, Gregoire:1992tf}.
For simple linear gauge choices, the anti-BRST invariance is an immediate consequence from the ordinary BRST symmetry and does not imply new physical content. 
Given a general gauge invariant theory, it is always possible to construct an anti-BRST/BRST invariant quantum action for a specific linear gauge condition \cite{Baulieu:1983tg}. However, for more intricate non-linear gauge conditions that is not always the case \cite{Hwang:1983sm, Baulieu:1983tg}.
Indeed, it has been shown for some specific situations that the anti-BRST plays an essential role to assure a consistent quantization, being a necessary condition for the decoupling of ghost states in the Hilbert physical subspace of the theory \cite{Baulieu:1983tg, Hwang:1989mn}.  Modern important instances of model quantizations with BRST and anti-BRST approaches as well as novel relevant applications of those symmetries appear often in the literature, from which we may cite the representative works \cite{Barcelos-Neto:1998nrq, Capri:2006cz, Kumar:2017sqz, Reshetnyak:2016evv, Capri:2020ppe, Dai:2020qpc, Thibes:2020yux,  Capri:2021pye, Kugo:2021bej, Varshovi:2020usc, Pandey:2020amp, Pandey:2020gfp, Berezhiani:2021zst, Mandal:2022xuw, Kugo:2022iob, Kugo:2022dui, Amaral:2022cje, Raval:2022eah}.
A nice complete review of anti-BRST transformations on a fair equal foot with its twin original BRST ones, for both Yang-Mills and more general gauge theories, focusing also on the generation of Ward identities, unitarity and renormalization issues, can be seen in reference \cite{Baulieu:1983tg}.
Modern reviews of the standard BFV quantization method and its connection with the BRST symmetry can be seen for instance in references \cite{Henneaux:1985kr, Teitelboim:1987rs, Henneaux:1992ig}.  Concerning a more direct Lagrangian approach, using BRST symmetry to discuss renormalizability and unitarity along the functional quantization framework we cite here the important concise review texts \cite{Piguet:1995er, Becchi:1996yh}.
On a more introductory level, the didactic review articles  \cite{Nemeschansky:1987xb, Niemi:1988bf, Zenteno:1994} are certainly also worth mentioning.

As it is clear from the above paragraph, nowadays the concept, importance, essence, role and applicability of the anti-BRST/BRST symmetries is very firmly well established.  Perhaps the same cannot yet be said about the so-called {\it dual} or {\it co}-BRST transformations \cite{Lavelle:1993xf, Tang:1994ru, Yang:1995nm, Yang:1995me, Malik:2000ty, Malik:1997ge} as well as other new similar BRST related symmetries or generalizations which have been more recently reported and discussed in the literature  \cite{Joglekar:1994tq,  Lahiri:2000ti, Rivelles:2001tq, Rai:2010aa, Rai:2010vu, Rai:2012yp, Dai:2020uab, Upadhyay:2019sgh, Mathieu:2021mxl, Rao:2022dra} and certainly deserve more profound studies.  From those we point up particularly the finite-field-dependent BRST transformations \cite{Joglekar:1994tq}  as a salient generalization which has shed light on a large amount of physical problems by connecting different gauge-fixings and Green's functions through field transformations in the partition function \cite{Rai:2012yp,   Upadhyay:2019sgh, Joglekar:1998dw, ff2,ff3,ff5,ff7,ff20,Deguchi:2016qge, Pandey:2018tmz, Mandal:2022xuw,Banerjee:2000jt}.  As for the dual-BRST symmetry, it first appeared as a somewhat ugly nonlocal and noncovariant symmetry in QED \cite{Lavelle:1993xf}, with such undesirable features being quickly relaxed in more polished versions in \cite{Tang:1994ru, Yang:1995nm}.  Nevertheless, it has been initially heavily criticized as being nothing more than a disguised ordinary BRST symmetry \cite{Rivelles:1995gb, Rabello:1995sa, Park:1995xj, Rivelles:1995hm}.  Incidentally, while studying cohomological aspects of the BRST symmetry, Yang and Lee associated a BRST adjoint generator to the Lavelle and McMullan dual-BRST symmetry \cite{Yang:1995me}, which finally allowed Malik and collaborators to house it, together with the ordinary BRST one, within a consistent Hodge theory framework on firm grounds \cite{Malik:1999qq, Malik:1999pb, Harikumar:2000ay, Malik:2000tx, Gupta:2008he, Gupta:2009bu, Gupta:2010epi, Kumar:2010kd, Krishna:2011xj, Kumar:2011if, Kumar:2012ur, Bhanja:2013vqa, Srinivas:2016iic, Kumar:2017qpe, Krishna:2018hqk}.  The BRST charges and operators characterize a de Rham cohomology structure in which a Hodge theory is well defined \cite{Ref0, Ref1, Ref2}.  Our goal in the present paper is to clarify the roles, similarities, differences, properties and further aspects of the mentioned four symmetries (BRST, anti-BRST, dual-BRST, anti-dual-BRST), an associated bosonic symmetry corresponding to the Laplacian operator in Hodge theory, as well as new sets of BRST symmetry transformations involving the Lagrange multiplier and ghost sectors within a concise and simple prototype gauge-invariant model, which can reproduce many other relevant ones previously approached in the literature within different contexts.  In this way, we isolate the main features of those symmetries in a model independent way -- our results are valid for all particular cases of the prototypical system.

Our prototypical system comes from the generalization of the model first presented in reference \cite{Thibes:2020yux}, in which we include an open potential function $V(q^k)$ enlarging its applicability by allowing additional dynamics.  This model can directly reproduce many others as particular cases, both mechanical \cite{Nemeschansky:1987xb, Gupta:2010epi, Shukla:2014spa, Bhanja:2015sha} and field theoretical ones \cite{Thibes:2020yux, Pandey:2021myh}, and also enjoys a strong similarity with other gauge field models \cite{Nemeschansky:1987xb}.  In fact, in reference \cite{Nemeschansky:1987xb} the rigid rotor was used to convey a deep analogy with QED and QCD, mainly regarding its (anti-)BRST symmetry aspects.  However, in \cite{Nemeschansky:1987xb} it was not so clear how a classical rigid rotor could be embodied with gauge symmetries.  That explanation in a systematic form for a generalized quantum rigid rotor can be seen in \cite{Thibes:2020yux}, where a construction of a gauge-invariant potential obtained along the Faddeev-Jackiw-BarcelosNeto-Wotzasek (FJBW) symplectic algorithm is shown.

In the present work, we scrutinize and organize all forms of BRST symmetries present on a prototypical first-class system, shedding light not just to many previous particular cases described by the present model but to various other possible similar field theory models.  In this way, we understand more clearly the similarities, differences and roles of the mentioned forms of BRST symmetries in a systematic way, being able to readily apply the obtained general results to particular cases of interest. In Section II below, we present a prototypical first-class system and perform its BFV quantization, calculating the Green's functions generating functional.  We also discuss aspects of its quantization on an extended Hilbert space including the ghost operators obtaining the conserved quantum BRST charge responsible for the ordinary BRST transformations.  In Section III, we obtain other forms of BRST transformations corresponding to the (anti-)(dual-)BRST symmetries of the quantum action.  We show how those symmetries can be obtained from each other by composing operations from a discrete group of symmetries of the action.  In Section IV, we show that the configuration space version of the previous various forms of BRST symmetries are endowed with a property which distinguishes the dual form, namely, while the BRST and anti-BRST transformations leave the classical action invariant, the dual-BRST and anti-dual-BRST ones leave the gauge-fixing term invariant.  In this sense, they complement each other.  In Section V, we discuss the algebra obtained from the BRST charge operators and obtain a corresponding bosonic symmetry associated to the Casimir operator of that algebra.  Section VI is reserved for the new BRST symmetries, a set of which have been recently discussed for different models in the literature and a brand new one.  We show that those new symmetries also have room within our prototypical system, accordingly being fully realized in the Lagrange multipliers and ghost fields sector.  We end in Section VII with our conclusion and final remarks.

\section{BFV Generating Functional and Ordinary BRST Symmetry}
We start by considering a prototypical first-class gauge-invariant dynamical system depending on the generalized canonical coordinates $(q^0,q^k,p_k)$, with $k=1,\dots,n$, defined by the Hamiltonian function
\begin{equation}\label{H}
H = U(q^k,p_k) + V(q^k)+q^0 T(q^k)\,,
\end{equation}
where $V(q^k)$ and $T(q^k)$ denote two given differentiable real functions.  The former represents an arbitrary physical potential, while the latter characterizes a first-class constraint condition imposed along the system dynamical evolution.
Still in equation (\ref{H}), we define further\footnote{Latin indexes run though $i,j,k,l=1,\dots,n$, with repeated index summation convention always assumed.} \cite{Thibes:2020yux}
\begin{equation}\label{U}
U(q^k,p_k)\equiv\frac{R^{ijkl}T_iT_jp_kp_l}{2f^{ij}T_iT_j}
\,,
\end{equation}
with $f^{ij}=f^{ij}(q^k)$ denoting a symmetric nondegenerated two-form,
$R^{ijkl}$ given by
\begin{equation}\label{Rijkl}
R^{ijkl}(q^m)\equiv f^{ij}f^{kl} - f^{ik}f^{jl}
\end{equation}
and $T_i$ standing for the $T$ partial derivative with respect to $q^i$, i.e.,
\begin{equation}
T_i\equiv\frac{\partial T}{\partial q^i}
\,.
\end{equation}
Note that $q^0$ appears in (\ref{H}) as a Lagrange multiplier enforcing the constraint relation $T(q^k)=0$ through its equation of motion.  For convenience, we treat it here as an extra coordinate with a corresponding canonical momentum $p_0$ comprising an additional constraint in phase space.
Although we define (\ref{H}) in principle as a mechanical system described by a discrete set of generalized coordinates depending on a time evolution parameter $t$, it can be easily extended to field theory models by considering (\ref{H}) and its following equations as written in DeWitt notation, in which the indexes $k$ and $0$ comprise also a continuous space dependence and the discrete summations include also spatial integrations.  
A thorough classical analysis of Hamiltonian (\ref{H}) can be found in reference \cite{Thibes:2020yux}, where it is shown that it indeed describes a first-class Hamiltonian dynamical system with applications to field theory models.  
The functional quantization of (\ref{H}) can be achieved, for instance, in terms of the Batalin-Fradkin-Vilkowsky (BFV) formalism \cite{Fradkin:1975cq, Batalin:1977pb, Fradkin:1977xi}, as we show in the following paragraphs.

According to the BFV quantization scheme, corresponding to the first class constraints $T(q^k)$ and $p^0$, we introduce a pair of Grassmann anticommuting ghosts variables (${\cal{C}},\bar{\cal{C}}$) along with their respective canonically conjugated momenta ($\bar{\cal{P}},\cal{P}$)
with ghost numbers gh ${\cal{C}}=$ gh ${\cal{P}}=1=-$gh ${\bar{\cal{C}}}=-$gh $\bar{\cal{P}}$, and
write the 
generating functional in the extended phase space as\footnote{We are using natural units, in which in particular $\hbar=1$.} 
\begin{equation}\label{Zpsi}
Z_\Psi =\int {\cal{D}}\varphi \; \exp (iS_{eff}) \,,
\end{equation}
where ${\cal{D}}\varphi$ represents the Liouville functional integration measure containing all variables of the theory, namely,
\begin{equation}
{\cal{D}}\varphi\equiv{\cal{D}}q^0\,{\cal{D}} p_0\,{\cal{D}}q^i\,{\cal{D}}p_i\,{\cal{D}}{{\cal{C}}}\,{\cal{D}}\bar{\cal{P}}\,{\cal{D}}{\bar{\cal{C}}}\,{\cal{D}}{\cal{P}}
\,.
\end{equation}
In equation (\ref{Zpsi}), $S_{eff}$ stands for the effective action
\begin{equation}
S_{eff}=\int dt\left ({\dot q}^i p_i + {\dot q}^0 p_0 + {\dot{\cal{C}}} \bar{\cal{P}} 
+{\dot{\bar{\cal{C}}}}{ \cal{P}} 
-H_\Psi \right )\,, \label{seff}
\end{equation}
which is written in terms of
the extended Hamiltonian $H_\Psi$ obtained from (\ref{H}) by adding the extra ghost variables and gauge-fixing parts as
\begin{equation}\label{Hpsi}
H_\Psi =  U(q^k,p_k) + V(q^k) + \left\{\Omega,\Psi\right\}
\,,
\end{equation}
with $\Omega$ denoting the BRST charge in the extended phase space generating the classical BRST symmetry transformations.
The gauge freedom is captured in the generating functional (\ref{Zpsi}) by the quantity $\Psi$ present in (\ref{Hpsi}), which denotes an arbitrary gauge fixing fermionic function of ghost number $-1$.
As physics should be gauge-indepentent, the Fradkin-Vilkowsky theorem \cite{Fradkin:1975cq, Batalin:1977pb,  Henneaux:1985kr} rightful assures the independence of (\ref{Zpsi}) from the particular choice of $\Psi$.

At quantum level, we promote all generalized coordinates from the extended phase space, including auxiliary and ghost variables, to linear operators acting on a Hilbert space satisfying (anti-)commutation relations directly obtained from the corresponding classical Poisson\footnote{Due to the BFV quantization scheme for first-class systems, we do not need Dirac brackets.} (anti-)brackets.  To be more precise, we write down below the non-null quantum fundamental relations\footnotemark[2]
\begin{equation}\label{QR}
\begin{array}{rllrll}
\left[ q^k, p_l \right]_- &=~ i\,\delta^k_{\,l} \,,&~~~~& \left[ \cal{C}, \bar{\cal{P}} \right]_+ &=~ -i \,,\\ 
\left[ q^0, p_0 \right]_- &=~ i \,,&~~~~&\left[ {\bar{\cal{C}}}, {{\cal{P}}} \right]_+ &=~ -i \,.
\end{array}
\end{equation}

The quantum BRST charge in the extended Hamiltonian phase space $\Omega_b$, the operatorial version of $\Omega$ in equation (\ref{Hpsi}), can be written as  \cite{Becchi:1974xu, Becchi:1974md, Tyutin:1975qk, Becchi:1975nq}
\begin{equation}
\Omega_b=i\left({\cal{C}}T(q^k)+ {{\cal{P}}}p_0 \right ) \label{Ob}
\,,
\end{equation}
and is responsible for generating the ordinary BRST symmetry.  In fact, given a generic function $F(q^k,p_k,q^0,p_0,{\cal C},\bar{\cal{P}} ,{\bar{\cal{C}}}, {{\cal{P}}})$ with well-defined Grassmann parity $\epsilon_F$, we define
\begin{equation}\label{deltaF}
\delta_b F={\left[F,\Omega_b\right]}_{\pm} \equiv F\Omega_b-(-1)^{\epsilon_F}\Omega_bF
\,.
\end{equation}
We shall often omit the $\pm$ subscript index when writing (anti-)commutation relations $\left[~~,~~\right]$ -- it should be always implied that we have an ordinary commutator when one of the arguments is bosonic (Grassmann-even) and an anticommutator when both arguments are fermionic (Grassmann-odd).
Note that, due to (\ref{QR}) represent the only non-null (anti-)commutation relations, there are no ordering ambiguities in (\ref{Ob}).
For the fundamental variables, plugging 
(\ref{Ob}) into (\ref{deltaF}) leads explicitly to
\begin{equation}\label{deltab}
\begin{array}{llll}
 \delta _b q^i =0\,,\quad \quad & \delta _b q^0 =-{\cal{P}}\,,\quad\quad & 
\delta _b {\cal{C}}=0\,,\quad\quad &
\delta _b
{\bar{\cal{C}}}=p_0 \,, \\
\delta _b p_i= {\cal{C}}T_i\,,\quad \quad & \delta _bp_0 =0\,,\quad \quad  & \delta _b {\bar{\cal{P}}}=T\,, \quad\quad
&  \delta _b {{\cal
{P}}}=0 \,.  
\end{array}
\end{equation}
The BRST charge $\Omega_b$ has ghost number $+1$ and is nillpotent, by which we mean
\begin{equation}
\Omega_b^2=0\,,
\end{equation}
and, as a direct consequence, the transformations (\ref{deltab}) are off-shell closed.
 
Following references \cite{Henneaux:1985kr, Henneaux:1992ig}, we
choose the gauge fixing fermion as 
\begin{equation}
\Psi= {\bar{\cal{P}}}q^0 +{\bar{\cal{C}}} \chi
\,,
\end{equation}
where
\begin{equation}\label{chi}
\chi\equiv\chi(q^k,p_k,q^0,p_0)
\end{equation}
is an arbitrary bosonic function which does not depend on the ghost variables.  Therefore, noting that
\begin{equation}
\left\{ \Omega_b, \Psi \right\} = q^0 T + {\cal{P}}{\bar{\cal{P}}}+i{\cal{C}}{\bar{\cal
{C}}}\left[T,\chi\right] + p_0 \chi +i{\cal P}\left[ p_0, \chi \right]
{\bar{\cal {C}}}
\,,
\end{equation}
from (\ref{Hpsi}) we may write the quantum BRST invariant Hamiltonian as
\begin{equation}\label{hatH}
\hat{H} = U + V +  q^0 T +  p_0 \chi +i{\cal{C}}\left[T,\chi\right] {\bar{\cal
{C}}} +i{\cal P}\left[ p_0, \chi \right]
{\bar{\cal {C}}} + {\cal{P}}{\bar{\cal{P}}}
\,.
\end{equation}
As the reader can check, regardless of the functional expression for $\chi$ in equation (\ref{chi}), the nillpotent ordinary BRST transformations (\ref{deltab})  leave the Hamiltonian (\ref{hatH}) invariant.

In order to have a nice handy working action and define a consistent framework for a later systematical discussion of various forms of BRST symmetries, particularly ready for comparison with previous studied cases in the literature, we consider next the standard gauge function
\begin{equation}\label{gf}
\chi=w^{-1}B+\frac{\xi}{2}p_0
\end{equation}
with the definitions
\begin{equation}
B=f^{ij}T_ip_j\,~~~\mbox{ and }~~~
w=f^{ij}T_iT_j\,.
\end{equation}
That specific standard form (\ref{gf}) leads to the extended gauge-fixed action
\begin{equation}\label{Sext}
S_{ext}=\int dt
\left (
{\dot q}^i p_i + {\dot{{\cal C}}}{{\bar{\cal{P}}}} + {\dot{\bar{\cal C}}}{{{\cal{P}}}} - U(q^k,p_k) - V(q^k) -q^0T(q^k) 
- \frac{\xi}{2}p_0^2 +p_0{\left( {\dot{q}}^0-w^{-1}B \right)}
+ {\cal C}{\bar{\cal{C}}} - {\cal P}{\bar{\cal{P}}}
\right )\,, 
\end{equation}
in which $\xi$, first introduced in (\ref{gf}), represents an open gauge-fixing parameter.  We may compare (\ref{gf}) to the $R_\xi$ gauges commonly used in QED and QCD, in which we have the limits of $\xi$ tending to $0$, $1$, $3$ and $\infty$ corresponding respectively to the Landau, Feynman-'t Hooft, Fried-Yennie and unitary gauges. The action (\ref{Sext}) is also invariant under the ordinary BRST transformations (\ref{deltab}).  In the next section, we shall introduce other similar BRST transformations representing new symmetries of action (\ref{Sext}).

\section{A Set of BRST-related Symmetries -- Hamiltonian Approach}
Still in the extended phase space, in a similar fashion to the ordinary BRST symmetry (\ref{deltab}) which we repeat below for clearness, we have other BRST-related symmetries enjoyed by action (\ref{Sext}).  If we take a careful look, we can see by direct inspection that the following four transformations, generated by their respective charges, leave the extended action (\ref{Sext}) invariant:
\subsection*{Ordinary BRST}
\begin{equation}  \tag{\ref{deltab}}
\begin{array}{llll}
 \delta _b q^i =0\,,\quad \quad & \delta _b q^0 =-{\cal{P}}\,,\quad\quad & 
\delta _b {\cal{C}}=0\,,\quad\quad &
\delta _b
{\bar{\cal{C}}}=p_0 \,, \\
\delta _b p_i= {\cal{C}}T_i\,,\quad \quad & \delta _bp_0 =0\,,\quad \quad  & \delta _b {\bar{\cal{P}}}=T\,, \quad\quad
&  \delta _b {{\cal
{P}}}=0 \,,
\end{array}
\end{equation}
\begin{equation}\label{Omegab}
\Omega_b=i\left[{\cal{C}}T(q^k)+ {{\cal{P}}}p_0 \right ]\,, ~~~~ \mbox{ gh }\Omega_b=+1\,.
\end{equation}

\subsection*{Anti-BRST}   
\begin{equation}\label{antideltab}
\begin{array}{llll}
 \bar{\delta} _b q^i =0\,,\quad \quad & \bar{\delta} _b q^0 =-{\bar{\cal{P}}}\,,\quad\quad & 
\bar{\delta} _b {\cal{C}}=p_0\,,\quad\quad &
\bar{\delta} _b
{\bar{\cal{C}}}=0 \,, \\
\bar{\delta} _b p_i= -{\bar{\cal{C}}}T_i\,,\quad \quad & \bar{\delta} _bp_0 =0\,,\quad \quad  & \bar{\delta} _b {\bar{\cal{P}}}=0\,, \quad\quad
&  \bar{\delta} _b {{\cal
{P}}}=-T \,,
\end{array}
\end{equation}
\begin{equation}\label{antiOmegab}
\bar{\Omega}_b=i\left[-{\bar{\cal{C}}}T(q^k)+ {\bar{\cal{P}}}p_0 \right ]\,, ~~~~ \mbox{ gh }\bar{\Omega}_b=-1\,.
\end{equation}

\subsection*{Dual-BRST} 
\begin{equation}\label{deltad}
\begin{array}{llll}
 {\bar{\delta}} _d q^i =0\,,\quad \quad & {\bar{\delta}} _d q^0 =-{\bar{\cal{C}}}\,,\quad\quad & 
{\bar{\delta}} _d {\cal{C}}=T\,,\quad\quad &
{\bar{\delta}} _d
{\bar{\cal{C}}}=0 \,, \\
{\bar{\delta}} _d p_i= {\bar{\cal{P}}}T_i\,,\quad \quad & {\bar{\delta}} _dp_0 =0\,,\quad \quad  & {\bar{\delta}} _b {\bar{\cal{P}}}=0\,, \quad\quad
&  {\bar{\delta}} _d {{
{\cal P}}}=p_0 \,,
\end{array}
\end{equation}
\begin{equation}\label{Omegad}
\bar{\Omega}_d=i\left[{\bar{\cal{P}}}T(q^k)+ {\bar{\cal{C}}}p_0 \right ]
\,, ~~~~ \mbox{ gh }\bar{\Omega}_d=-1\,.
\end{equation}

\subsection*{Anti-dual BRST}  
\begin{equation}\label{antideltad}
\begin{array}{llll}
 {\delta} _d q^i =0\,,\quad \quad & {\delta} _d q^0 =-{{\cal{C}}}\,,\quad\quad & 
{\delta} _d {\cal{C}}=0\,,\quad\quad &
{\delta} _d
{{\bar{\cal{C}}}}=-T \,, \\
{\delta} _d p_i= -{{\cal{P}}}T_i\,,\quad \quad & {\delta} _dp_0 =0\,,\quad \quad  & {\delta} _d {{\bar{\cal{P}}}}=p_0\,, \quad\quad
&  {\delta} _d {{\cal
{P}}}=0 \,,
\end{array}
\end{equation}
\begin{equation}\label{antiOmegad}
{\Omega}_d=i\left[-{{\cal{P}}}T(q^k)+ {{\cal{C}}}p_0 \right ]
\,, ~~~~ \mbox{ gh }{\Omega}_d=+1\,.
\end{equation}
All charges introduced above in equations (\ref{Omegab}), (\ref{antiOmegab}), (\ref{Omegad}) and (\ref{antiOmegad}) are off-shell nillpotent and, consequently, the corresponding symmetries are off-shell closed.
It is worth noticing that the two dual symmetries, ${\bar{\delta}}_d$ and ${\delta}_d$, can be obtained respectively from $\delta_b$ and $\bar{\delta_b}$ by the shifting
\begin{equation}\label{can1}
a:~~~{\cal{C}} \longrightarrow {\bar{\cal{P}}} \,, ~~~ {\bar{\cal{C}}} \longrightarrow {{\cal{P}}} \,, ~~~ {{\cal{P}}} \longrightarrow {\bar{\cal{C}}}
\,, ~~~   {\bar{\cal{P}}}  \longrightarrow
{\cal{C}}
\,,~~~~\mbox{ (anti-)BRST }\longrightarrow\mbox{ (anti-)dual-BRST }\,,
\end{equation}
while the anti- symmetries, $\bar{\delta}_b$ and ${\delta}_d$, can be generated respectively from  $\delta_b$ and ${\bar{\delta}}_d$ by
\begin{equation}\label{can2}
b:~~~{\cal{C}} \longrightarrow  - {\bar{\cal{C}}} \,, ~~~ {\bar{\cal{C}}} \longrightarrow  {\cal{C}}   \,, ~~~ {{\cal{P}}} \longrightarrow {\bar{\cal{P}}} 
\,, ~~~  {\bar{\cal{P}}} \longrightarrow
- {{\cal{P}}} 
\,,~~~~\mbox{ (dual-)BRST }\longrightarrow\mbox{ (dual-)anti-BRST }\,.
\end{equation}
Both equations (\ref{can1}) and (\ref{can2}) above represent canonical transformations in the sense of preserving the fundamental relations (\ref{QR}).

What is the origin of all those similar, seemingly related, symmetries?
It turns out that action (\ref{Sext}) is invariant with respect to the group $\mathbb{Z}_4\times\mathbb{Z}_2$ of discrete permutations among the ghost variables.  As a matter of fact, the transformations (\ref{can1}) and (\ref{can2}) can be concretely taken as the two generators, respectively $a$ and $b$, for the $\mathbb{Z}_4\times\mathbb{Z}_2$ group presentation given by
\begin{equation}
\mathbb{Z}_4\times\mathbb{Z}_2 = \,\, <a,b \,|\, a^2=b^4=e, ab=ba> \,.
\end{equation}
That is a symmetry group of order eight, whose elements give rise to the transformations $\pm$BRST, $\pm$anti-BRST, $\pm$dual-BRST, $\pm$anti-dual-BRST\footnote{ Please note that if we multiply the  Eqs \ref {deltab}, \ref{antideltab}. \ref{deltad},\ref{antideltad} by $(-1)$ we get additional trivial symmetries.} Of course the same happens with any BRST-invariant action, the corresponding ghost permutation group of symmetries of the action produces other BRST-related symmetries.  This is one of the reasons for which the dual-BRST symmetry was initially criticized on the literature -- in a sense, the BRST-related symmetries (\ref{antideltab}), (\ref{deltad}) and (\ref{antideltad}) come simply from a relabeling of ghost variables in (\ref{deltab}) and do not produce new physics.  That may be true in a Hamiltonian context as we see here, however, in a gauge-fixed Lagrangian context, the dual-BRST symmetry indeed assumes a prominent distinct role as we shall soon see. 

We stress once more that all previous symmetries occur in the extended phase space, well-suited for a Hamiltonian approach.  Nevertheless, their original discovery can be clearly retrieved on the past literature within somewhat different contexts, namely, ordinary BRST \cite{Becchi:1974xu, Becchi:1974md, Tyutin:1975qk, Becchi:1975nq}, anti-BRST \cite{Curci:1976bt, Curci:1976ar, Ojima:1980da} and dual-BRST \cite{Lavelle:1993xf, Tang:1994ru, Yang:1995nm, Yang:1995me, Malik:2000ty, Malik:1997ge}.   
Here, we see all of them interconnected within a systematic unified treatment.
Usually, genuine dual-BRST transformations are characterized by the invariance of the gauge-fixing term.  In the present case however,  by defining
\begin{equation}\label{Sgf}
S_{gf}\equiv
\int dt
\left (
  p_0{\left( {\dot{q}}^0-w^{-1}B \right)}   - \frac{\xi}{2}p_0^2 
\right)
\end{equation}
we have
\begin{equation}
{\bar{\delta}}_d \, S_{gf}=-\int dt \, p_0 \left(
\dot{\bar{\cal C}}+\bar{\cal P} \right)
\,,
\end{equation}
and similarly
\begin{equation}
\delta_d \, S_{gf}= \int dt \, p_0 \left( {\cal P} - \dot{\cal C} \right)
\,.
\end{equation}
In fact,
in order to have the well-known elegant interpretation corresponding to the plain invariances of the classical action and gauge-fixing terms respectively under (anti-)BRST and (anti-)dual-BRST transformations, we have to come closer to a Lagrangian form.  That can be done by momenta functional integrations, as we show in the next section.

\section{BRST Symmetries in Configuration Space}
In this section, we investigate further the various BRST symmetries of the prototypical system (\ref{H}) closer to a gauge-fixed Lagrangian viewpoint.
With that goal in mind, let us consider the possibility of functional integration of (\ref{Zpsi}) in the momenta variables $p_k,p_0, {\cal P}$ and ${\bar{\cal{P}}}$.  Since we are considering a generic constraint function $T(q^k)$, which reflects itself on the functional form of the potential (\ref{U}), it is not advisable to perform the $p_k$ integration.  Actually it is of good measure to maintain $p_k$ in the theory, as its existence keeps the gauge symmetry simple \cite{Thibes:2020yux} and we can interpret the system as defined by a first-order Lagrangian \cite{Amorim:1995sh}.   Concerning the momenta variable $p_0$, we see that it appears in the action (\ref{Sext}) only through the gauge-fixing term $S_{gf}$ defined in (\ref{Sgf}).  Performing a functional integration over $p_0$ would bring down an expression of the form
\begin{equation}
\frac{1}{2\xi}{\left( {\dot{q}}^0-w^{-1}B \right)}^2
\end{equation}
which, as argued by the end Section II, can be compared to the usual quadratic covariant gauge-fixing term present in the QED and QCD quantum Lagrangians.  However, as it is well-known, the presence of this term spoils the off-shell nilpotency of the BRST symmetries which become closed only by means of the equations of motion, i.e., on shell.  In this way, we see clearly that $p_0$ plays the role of a Nakanishi-Lautrup variable \cite{Nakanishi:1966zz, Lautrup:1967zz} and should remain alive on the theory for the sake of off-shell BRST closure.
Therefore, we perform the functional integration of (\ref{Zpsi}) only in the ghost momenta variables $\cal{P}$ and $\bar{\cal{P}}$, after which we obtain a neat first-order action in the exponential argument of $Z_\Psi$ given by
\begin{equation}\label{S}
S_{}=\int dt
\left (
{\dot q}^i p_i - {\dot{{\cal C}}}{\dot{\bar{\cal{C}}}} - U(q^k,p_k) - V(q^k) -q^0T(q^k) 
- \frac{\xi}{2}p_0^2 +p_0{\left( {\dot{q}}^0-w^{-1}B \right)}
+ {\cal C}{\bar{\cal{C}}}
\right )\,.
\end{equation}
To that extent, after the $\cal{P}$ and $\bar{\cal{P}}$ functional integrations, the new symmetries and respective conserved charges of the new concise action (\ref{S}) corresponding to equations (\ref{deltab}) and (\ref{Omegab}) to (\ref{antiOmegad}) become now
\subsection*{Ordinary BRST}
\begin{equation}\label{sb}
\begin{array}{lll}
 s _b\, q^i =0\,,\quad \quad & s _b\, q^0 =-{\dot{\cal{C}}}\,,\quad\quad & 
s _b\, {\cal{C}}=0\,, \\
s _b\, p_i= {\cal{C}}T_i\,,\quad \quad & s _b\,p_0 =0\,,\quad \quad  & s _b\, {\bar{\cal{C}}}=p_0  \,,
\end{array}
\end{equation}
\begin{equation}\label{Qb}
{\cal Q}_b=i\left[{\cal{C}}T(q^k)+ {{\dot{\cal{C}}}}p_0 \right ] \,, ~~~~ \mbox{ gh }{\cal Q}_b=+1 \,;
\end{equation}
\subsection*{Anti-BRST}
\begin{equation}\label{antisb}
\begin{array}{lll}
 \bar{s} _b\, q^i =0\,,\quad \quad & \bar{s} _b\, q^0 ={\dot{\bar{\cal{C}}}}\,,\quad\quad & 
\bar{s} _b\, {\cal{C}}=p_0\,, \\
\bar{s} _b\, p_i= -{\bar{\cal{C}}}T_i\,,\quad \quad & \bar{s} _b\,p_0 =0\,,\quad \quad  & \bar{s} _b\, {\bar{\cal{C}}}=0  \,,
\end{array}
\end{equation}
\begin{equation}\label{antiQb}
\bar{{\cal Q}}_b=i\left[-{\bar{\cal{C}}}T(q^k) - {\dot{\bar{\cal{C}}}}p_0 \right ]
\,, ~~~~ \mbox{ gh }\bar{{\cal Q}}_b=-1\,;
\end{equation}
\subsection*{Dual-BRST}
\begin{equation}\label{sd}
\begin{array}{lll}
 \bar{s} _d\, q^i =0\,,\quad \quad & \bar{s} _d\, q^0 =-{\bar{\cal{C}}}\,,\quad\quad & 
\bar{s} _d\, {\cal{C}}=T\,, \\
\bar{s} _d\, p_i= -{\dot{\bar{\cal{C}}}}T_i\,,\quad \quad & \bar{s} _d\,p_0 =0\,,\quad \quad  & \bar{s} _d\, {\bar{\cal{C}}}=0  \,,  
\end{array}
\end{equation}
\begin{equation}\label{Qd}
{\bar{{\cal Q}}}_d=i\left[-{\dot{\bar{\cal{C}}}}T(q^k)+ {\bar{\cal{C}}}p_0 \right ]
 \,, ~~~~ \mbox{ gh }{\bar{{\cal Q}}}_d=-1 \,;
\end{equation}
\subsection*{Anti-Dual-BRST}
\begin{equation}\label{antisd}
\begin{array}{lll}
 {s} _d\, q^i =0\,,\quad \quad & {s} _d\, q^0 =-{{\cal{C}}}\,,\quad\quad & 
{s} _d\, {\cal{C}}=0\,, \\
{s} _d\, p_i= -{\dot{{\cal{C}}}}T_i\,,\quad \quad & {s} _d\,p_0 =0\,,\quad \quad  & {s} _d\, {\bar{\cal{C}}}=-T  \,,
\end{array}
\end{equation}
\begin{equation}\label{antiQd}
{{\cal Q}}_d=i\left[-{\dot{\cal{C}}}T(q^k)+ {{\cal{C}}}p_0 \right ]
\,, ~~~~ \mbox{ gh }{{\cal Q}}_d=+1\,.
\end{equation}
All the above transformations are off-shell nillpotent and each one of them can be checked to characterize a particular symmetry of the action (\ref{S}).  Furthermore, now the two dual-BRST symmetries do leave the gauge-fixing term invariant.

Indeed,
it is interesting to see at this point that, while the ordinary BRST and anti-BRST transformations leave the original classical action invariant,\footnote{Recall that the usual (anti-)BRST transformations can be obtained from the gauge symmetry of the classical action by replacing the gauge parameters by the (anti-)ghosts.} with the variation of the gauge-fixing term being canceled by the necessary ghost terms, on the other hand, the dual and anti-dual BRST transformations leave the gauge-fixing term invariant by itself, with a mutual cancellation between the variations coming from the classical Lagrangian and ghost terms.  This last reasoning may be simpler expressed in symbols as
\begin{equation}
s_b \, S_c = {\bar{s}}_b \, S_c = 0
\end{equation}
and
\begin{equation}
{\bar{s}}_d \, S_{gf} = {{s}}_d \, S_{gf} = 0
\end{equation}
with
\begin{equation}
S_c \equiv
\int dt
\bigg\lbrace
{\dot q}^i p_i  - U(q^k,p_k) - V(q^k) -q^0T(q^k) 
\bigg\rbrace
\end{equation} representing the classical action
and the ghost-fixing term $S_{gf}$ given by equation (\ref{Sgf}).  Along this line, in BRST-cohomology terms, we understand the BRST invariances of the action $S$ in the sense that it can be decomposed as a sum between a BRST-exact and a BRST-closed parts.  This assertion holds for each one of the four BRST symmetries, including the dual ones, as we may conveniently rewrite (\ref{S}) in one of the two equivalent forms below
\begin{equation}
S_{}=\int dt
\bigg\lbrace
{\dot q}^i p_i  - U(q^k,p_k) - V(q^k) -q^0T(q^k) + \frac{1}{2}s_b\,\bar{s}_b\ \big[ \xi {\cal C}{\bar{\cal C}} - {(q^0)}^2 +w^{-2}B^2 \big]
\bigg\rbrace
\end{equation}
or
\begin{equation}
S_{}=\int dt
\bigg\lbrace
- U(q^k,p_k) - V(q^k) - \frac{\xi}{2}p_0^2 +p_0{\left( {\dot{q}}^0-w^{-1}B \right)}
+ \frac{1}{2} s_d\, \bar{s}_d\,  {
\Big[
\left(      \frac{  {\dot q}^i p_i }{\dot{T}} \right)}^2
-(q^0)^2 
\Big]
\bigg\rbrace
\,.
\end{equation}
Therefore, we have seen a first important feature which distinghishes the (anti-)BRST from the (anti-)dual-BRST symmetries.  Additionally, the current set of symmetries described in equations (\ref{sb}) to (\ref{antisd}) allows for a physical realization of a Hodge theory \cite{Ref0, Ref1, Ref2} with a rich algebraic structure, as shall be clear in the next section.

\section{BRST Algebra}
In this section we investigate the Lie superalgebra generated by the previous BRST charges.  We start by recalling that all four introduced BRST charges (\ref{Qb}), (\ref{antiQb}), (\ref{Qd}) and (\ref{antiQd}) are fermionic operators and fully off-shell nillpotent, i.e.,
\begin{equation}\label{Q2=0}
{\cal Q}_b^2=\bar{{\cal Q}}_b^2=\bar{{\cal Q}}_d^2={\cal Q}_d^2=0\,.
\end{equation}
Since we are at quantum level, the properties (\ref{Q2=0}) are not trivial, but rather assure the closure of the corresponding symmetry transformations.  
Besides satisfying property (\ref{Q2=0}), we stress that all BRST charges are conserved under time evolution modulo equations of motion.
We may additionally introduce a ghost number operator given by
\begin{equation}\label{GN}
{\cal G} = i \left( {\dot{{\cal{C}}}}  {\bar{\cal{C}}} - {\cal C} {\dot{\bar{\cal{C}}}} \right)\,,
\end{equation}
satisfying
\begin{equation}
\begin{aligned}
\left[ {\cal G} ,  {\cal Q}_b \right] = {\cal Q}_b\,,~~~~~\left[ {\cal G} ,  \bar{{\cal Q}}_b \right] =  -\bar{{\cal Q}}_b \,, \\
\left[ {\cal G} ,  {\bar {\cal Q}}_d \right] = -{\bar {\cal Q}}_d\,, ~~~~~ [ {\cal G} ,  {\cal Q}_d ] = {\cal Q}_d \,,
\end{aligned}
\end{equation}
which confirms the charges ghost numbers.
Ghost number conservation is then warranted by the global scale symmetry
\begin{equation}\label{scale}
{{\cal{C}}}  \longrightarrow  e^\lambda {{\cal{C}}}\,,~~~~~~
{\bar{\cal{C}}}  \longrightarrow e^{-\lambda} {\bar{\cal{C}}}\,,
\end{equation}
with $\lambda$ denoting a continuous constant parameter.  Indeed, transformations (\ref{scale}) clearly leave the quantum action (\ref{S}) invariant, while time conservation of (\ref{GN}) follows directly from the equations of motion.

The two (anti-)BRST operators commute among themselves
\begin{equation}
\left[{\cal Q}_b,\bar{{\cal Q}}_b\right]= 0
\,,
\end{equation}
as well as the (anti-)dual-BRST ones
\begin{equation}
\left[\bar{{\cal Q}}_d,  {\cal Q}_d \right] = 0
\,,
\end{equation}
while we have a non-null anticommutator between the (anti-)BRST and (anti-)dual-BRST given by
\begin{equation}\label{W}
\left[  {{\cal Q}}_b  ,  \bar{{\cal Q}}_d  \right]=
\left[  \bar{{\cal Q}}_b  ,  {{\cal Q}}_d  \right]=
i(T^2+p_0^2) \equiv 2 {\cal W}
\,.
\end{equation}
The above defined bosonic quantity ${\cal W}$ represents a Casimir operator for the superalgebra generated by the BRST charges, as it commutes with all of them, and gives rise to a new bosonic transformation defined for an arbitrary function $F$ as
\begin{equation}
s_W F \equiv [ F, {\cal W} ]
\,. 
\end{equation}
For the fundamental variables of the theory, the non-null $s_W$ tranformations read explicitly
\begin{equation}
s_W p_i = T T_i\,,~~~~~~
s_W q^0 = -p_0\,,
\end{equation}
and constitute a new symmetry leaving (\ref{S}) invariant.  Similarly to the BRST charges, the Casimir operator $\cal W$ is a constant of motion being conserved along time evolution modulo equations of motion, albeit being a bosonic quantity.

\section{New Symmetries}
In none of the previously seen symmetries does the Nakanishi-Lautrup variable $p_0$ transform.  In this last section we address the realization of the new recent symmetries involving the Lagrange multiplier, Nakanishi-Lautrup and ghosts sector in the current prototypical first-class system.   Inspired by the fact that the gauge-fixing term (\ref{Sgf}) is left invariant under the substitution
\begin{equation}
p_0 \rightarrow
-p_0 +\frac{2({\dot{q}}^0-w^{-1}B)}{\xi}
\,,
\end{equation}
it is possible to look for correspondingly new forms of BRST transformations involving $p_0$.  Accordingly,
in the same fashion as the particular models discussed in references \cite{Lahiri:2000ti, Rivelles:2001tq, Rai:2010aa}, the quantum action (\ref{S}) enjoys the further nillpotent symmetries
\begin{equation}
\Delta_1 p_i =  {\cal C}T_i\,,
\end{equation}
\begin{equation}
\Delta_1 p_0 = -\frac{2}{\xi}({\cal C}+\ddot{\cal C})\,,
\end{equation}
\begin{equation}
\Delta_1 q^0 = - \dot{\cal C}\,,
\end{equation}
\begin{equation}
\Delta_1 \bar{\cal C} =
- p_0+\frac{2}{\xi} {\left( {\dot{q}}^0-w^{-1}B \right)} 
\,,
\end{equation}
of ghost number $1$
and its corresponding anti mirror   
\begin{equation}
\bar{\Delta}_1 p_i = - \bar{\cal C}T_i
\,,
\end{equation}
\begin{equation}
\bar{\Delta}_1 p_0 = \frac{2}{\xi}(\bar{\cal C}+\ddot{\bar{\cal C}})
\,,
\end{equation}
\begin{equation}
\bar{\Delta}_1 q^0 = \dot{\bar{\cal C}}
\,,
\end{equation}
\begin{equation}
\bar{\Delta}_1 {\cal C} =
 p_0-\frac{2}{\xi} {\left( {\dot{q}}^0-w^{-1}B \right)} 
\,,
\end{equation}
of ghost number $-1$.

Additionally, we report here a brand new set of non-local symmetries for action (\ref{S}) with ghost numbers $1$ and $-1$ given, respectively, by
\begin{equation}
\Delta_2 p_0 = \frac{1}{\xi}({\cal C}+\ddot{\cal C})
\,,
\end{equation}
\begin{equation}
\Delta_2 q^0 = {1\over2}\dot{\cal C}+{1\over2}\int {\cal C} \,dt
\,,
\end{equation}
\begin{equation}
\Delta_2 \bar{\cal C} =
{1\over2}p_0-\frac{1}{\xi}({\dot{q}}^0-w^{-1}B)
-{1\over2}\int T\,dt
\,,
\end{equation}
and
\begin{equation}
\bar{\Delta}_2 p_0 = -\frac{1}{\xi}(\bar{\cal C}+\ddot{\bar{\cal C}})
\,,
\end{equation}
\begin{equation}
\bar{\Delta}_2 q^0 = -{1\over2}\dot{\bar{{\cal C}}}-{1\over2}\int {\bar{\cal C}} \,dt
\,,
\end{equation}
\begin{equation}
\bar{\Delta}_2 {\cal C} =
{1\over2}p_0-\frac{1}{\xi}({\dot{q}}^0-w^{-1}B)
-{1\over2}\int T\,dt
\,.
\end{equation}
The above $\Delta_2$ symmetries are clearly distinct from the $\Delta_1$ ones, as the former do not affect $p_i$, i.e.,
\begin{equation}
\Delta_2 p_i =\bar{\Delta}_2 p_i =0\,,
\end{equation}
while the latter contains terms corresponding to integrals of $T$ and $\cal C$.
A comparative analysis of the $\Delta$ symmetries as well as its relevance in specific quantum field theory models is currently under investigation.

\section{Conclusion}
The quantization of the prototypical first-class system introduced in Section II along the functional BFV procedure has allowed us to realize a considerable set of forms of BRST transformations constituting symmetries at quantum level, which we then deeply analyzed in detail.  Those symmetries comprised not just the ordinary BRST ones but included also the dual or co-BRST ones which have appeared in the literature in a plethora of different field theory models in many different contexts.   By clarifying the action of a discrete group of transformations on the ghosts sector, we have seen that the dual and anti BRST symmetries can be freely interchanged among a total of eight possibilities connected by canonical transformations in a Hamiltonian framework.  When coming to a gauge-fixed Lagrangian approach, the dual BRST symmetries achieve its full glorious interpretation, displaying its main characteristic of leaving the gauge-fixing term invariant -- a dual behavior when compared to the ordinary BRST symmetry which leaves instead the classical action invariant.  Further, the BRST charges exhibit the Hodge theory properties and it is possible to define a further Casimir operator leading to an extra bosonic symmetry and closing a Lie superalgebra among the conserved symmetry generators.
The simplicity and generality of the chosen first-class prototypical system permits those interpretations to be extended to other similar models in the literature shedding light on previous controversies regarding the physical interpretation of the dual BRST symmetries.  In a similar fashion to some of the mentioned field theory models, the prototypical first-class system also exhibits new BRST symmetries involving the Lagrange multipliers, Nakanish-Lautrup variable and ghosts.  Additionally, we have had the opportunity to report brand new forms of BRST symmetries fully realized in the prototypical first-class system.

{\bf Acknowledgements:} BPM acknowledges the research Grant for faculty under IoE Scheme (Number 6031) of Banaras Hindu University, Varanasi.

\end{document}